\documentclass[final,12pt,1p,times]{article}

\usepackage{graphicx}
\def\ben{\begin{equation}}
\def\een{\end{equation}}

\let\a=\alpha   \let\d=\delta 
    
  \let\n=\nu   
 \let\t=\tau    
       \let\D=\Delta  
     
\let\C=\Chi

\def\nn{\nonumber} \def\bd{\begin{document}} \def\ed{\end{document}}
\def\ds{\documentstyle} \let\fr=\frac \let\bl=\bigl \let\br=\bigr
\let\Br=\Bigr \let\Bl=\Bigl
\let\bm=\bibitem
\let\na=\nabla
\let\pa=\partial \let\ov=\overline
\newcommand{\be}{\begin{equation}}
\newcommand{\ee}{\end{equation}}
\def\ba{\begin{array}}
\def\ea{\end{array}}
\def\ft#1#2{{\textstyle{{\scriptstyle #1}\over {\scriptstyle #2}}}}
\def\fft#1#2{{#1 \over #2}}
\def\del{\partial}
\def\vp{\varphi}
\def\sst#1{{\scriptscriptstyle #1}}
\def\oneone{\rlap 1\mkern4mu{\rm l}}
\def\td{\tilde}
\def\wtd{\widetilde}
\def\ie{\rm i.e.\ }
\def\dalemb#1#2{{\vbox{\hrule height .#2pt
        \hbox{\vrule width.#2pt height#1pt \kern#1pt
                \vrule width.#2pt}
        \hrule height.#2pt}}}
\def\square{\mathord{\dalemb{6.8}{7}\hbox{\hskip1pt}}}
\newcommand{\ho}[1]{$\, ^{#1}$}
\newcommand{\hoch}[1]{$\, ^{#1}$}
\newcommand{\bea}{\begin{eqnarray}}
\newcommand{\eea}{\end{eqnarray}}
\newcommand{\ra}{\rightarrow}
\newcommand{\lra}{\longrightarrow}
\newcommand{\Lra}{\Leftrightarrow}
\newcommand{\bp}{\tilde \beta^\prime}
\newcommand{\tr}{{\rm tr} }
\newcommand{\Tr}{{\rm Tr} }
\def\0{{\sst{(0)}}}
\def\1{{\sst{(1)}}}
\def\2{{\sst{(2)}}}
\def\3{{\sst{(3)}}}
\def\4{{\sst{(4)}}}
\def\5{{\sst{(5)}}}
\def\6{{\sst{(6)}}}
\def\7{{\sst{(7)}}}
\def\8{{\sst{(8)}}}
\def\n{{\sst{(n)}}}
\def\cA{{{\cal A}}}
\def\cB{{{\cal B}}}
\def\cF{{{\cal F}}}
\def\cH{{{\cal H}}}
\def\tV{\widetilde V}
\def\tW{\widetilde W}
\def\tH{\widetilde H}
\def\tE{\widetilde E}
\def\tF{\widetilde F}
\def\tA{\widetilde A}
\def\im{{i}}
\def\tY{{{\wtd Y}}}
\def\ep{{\epsilon}}
\def\vep{{\varepsilon}}
\def\R{\rlap{\rm I}\mkern3mu{\rm R}}
\def\bD{{{\bar D}}}

\def\R{\rlap{\rm I}\mkern3mu{\rm R}}
\def\bD{{{\bar D}}}
\def\R{{{\Bbb R}}}
\def\C{{{\Bbb C}}}
\def\H{{{\Bbb H}}}
\def\CP{{{\Bbb C}{\Bbb P}}}
\def\RP{{{\Bbb R}{\Bbb P}}}
\def\Z{{{\Bbb Z}}}
\def\bA{{{\Bbb A}}}
\def\bB{{{\Bbb B}}}
\def\bC{{{\Bbb C}}}
\def\bD{{{\Bbb D}}}
\def\bE{{{\Bbb E}}}
\def\bZ{{{\Bbb Z}}}
\def\Re{{{\frak{Re}}}}
\def\Im{{{\frak{Im}}}}
\def\cosec{{\,\hbox{cosec}\,}}
\def\Gm{{\Gamma_{\!\! -}}}
\def\Gp{{\Gamma_{\!\! +}}}
\def\stan{{standard }}
\def\nonstan{{supernumerary }}

\newcommand{\tamphys}{\it Center for Theoretical Physics,
Texas A\&M University, College Station, TX 77843}

\newcommand{\upenn}{\it Department of Physics and Astronomy,\\ University
of Pennsylvania, Philadelphia, PA 19104}

\newcommand{\brussels}{\it Physique Th\'eorique et Math\'ematique,
Universit\'e Libre de Bruxelles,\\ Campus Plaine C.P. 231, B-1050
Bruxelles, Belgium}

\newcommand{\auth}{Zhiwei Chong \hoch{1} and Yajun Wei \hoch{2}}

\begin{document}

\begin{flushright}
\today
\end{flushright}

\vspace{10pt}

\begin{center}

{\large {\bf  Novel Approaches to Solve Simple Harmonic Motion}}

\vspace{20pt}
\auth

\vspace{10pt}{\footnote{chong.zhiwei@gmail.com} \it International Division, Experimental School Affiliated with Zhuhai No.1 High School, Zhuhai, Guangdong, China}\\
\vspace{10pt}{ \footnote{runnerwei@qq.com} \it Zhuhai No.1 High School, Zhuhai, Guangdong, China}



%
%
%

\vspace{20pt}


\begin{abstract}
This paper presents two novel approaches to solve the classic simple harmonic motion. In one approach, the distance between the equilibrium position and the maximal displacement is divided into~$N$ equal segments. In each segment, the mass moves with constant acceleration under the average of two forces at the ends of the segment. Summing up the time covering each segment and taking the large-$N$ limit reproduce one quarter of the period for simple harmonic motion. In the other approach, the time moving from the maximal displacement to the equilibrium position is divided into~$N$ equal intervals. A recurrence relation for the displacement is obtained. The large-$N$ limit of its solution results in the same solution as that obtained from solving differential equation.


\end{abstract}

\end{center}



\section{Introduction}
Simple harmonic motion (henceforth SHM, with a spring-mass system in mind) is treated in all introductory physics textbooks \cite{giancoli, sears, kleppner, morin} and at least one calculus textbook~\cite{thomas}. Other than the approaches in these textbooks, this paper presents two new approaches which discretize the  displacement (time) into $N$ segments (intervals) and take the large-$N$ limit to obtain the solution. 
They are conceptually easy but technically slightly challenging. Nevertheless, the relevant mathematics is still within the reach of most undergraduate students or even good high school students. 

In textbooks, mainly there are three approaches to treat SHM. Firstly, a connection is established between SHM and the projection of circular motion on a diameter \cite{giancoli, sears}.\footnote{
William Thomson (also known as Lord Kelvin) and Peter Tait coined the term ``simple harmonic motion" in~1867~\cite{treatise}. It is Newton who is the first to use a circle to describe SHM. One derived property of this motion is that its acceleration is proportional to its displacement, but they are in opposite directions. This property later became the definition of SHM in nowadays physics textbooks.
}
The merit of this approach is that it is accessible to students without knowledge in calculus. 
One drawback , however, is its lack of a dynamical foundation in the sense that there is no force generating the  motion. Thereby, it is criticized as nonphysical, arbitrary,
artificial, imaginary, and even confusing to students \cite{treatment, noncalculus, novel, introducing}.
Secondly, the solution to the differential equation is presented, and students are invited to verify the solution~\cite{giancoli,kleppner}. This approach does not involve solving differential equation but only requires students to have some experience in differentiation. Clearly this half-baked approach is not satisfactory but a mere convenience.
Lastly, the differential equation is solved either by integrating twice~\cite{kleppner} or using techniques in the theory of differential equations \cite{morin, thomas}. 
The contribution of this paper lies in that it \emph{solves} a classic problem in novel ways on one hand with a limited amount of mathematical preparation on the other. 

Section \ref{prelim} illustrates the basic idea of this paper. The method is developed systematically in 
Sec.~\ref{riemann}. The total displacement is divided into $N$ equal segments. In each segment, the mass moves with constant acceleration under the average of two elastic forces at each end of the segment under consideration. The total time is obtained as the large-$N$ limit of the sum of time covering each segment.
Section~\ref{recursive} presents another approach in which the time instead is divided into $N$ equal intervals. A recurrence relation for the displacement is obtained. Taking the large-$N$ limit of its solution yields the same solution as that from solving differential equation. We conclude and discuss in Sec. \ref{con}.

\section{Preliminary Consideration}
\label{prelim}

Suppose we have not learned calculus,  in particular differential equations, but we still want to obtain the period for the SHM with whatever mathematics we have at hand. We do not expect too much, and we persuade ourselves to be satisfied with an approximate solution for the moment. But we aim to obtain a value that is as accurate as we can. Note this is not a numerical method paper, and we will develop this approach systematically to obtain the exact solution.

A spring-mass system consists of a mass $m$ and a massless spring with spring constant~$k$. One end of the spring is fixed and the other is attached with a mass $m$ which lies on a frictionless horizontal surface, as is illustrated in Fig. 1.
\begin{figure}[h!]
\centering
\includegraphics[width=4.0in]{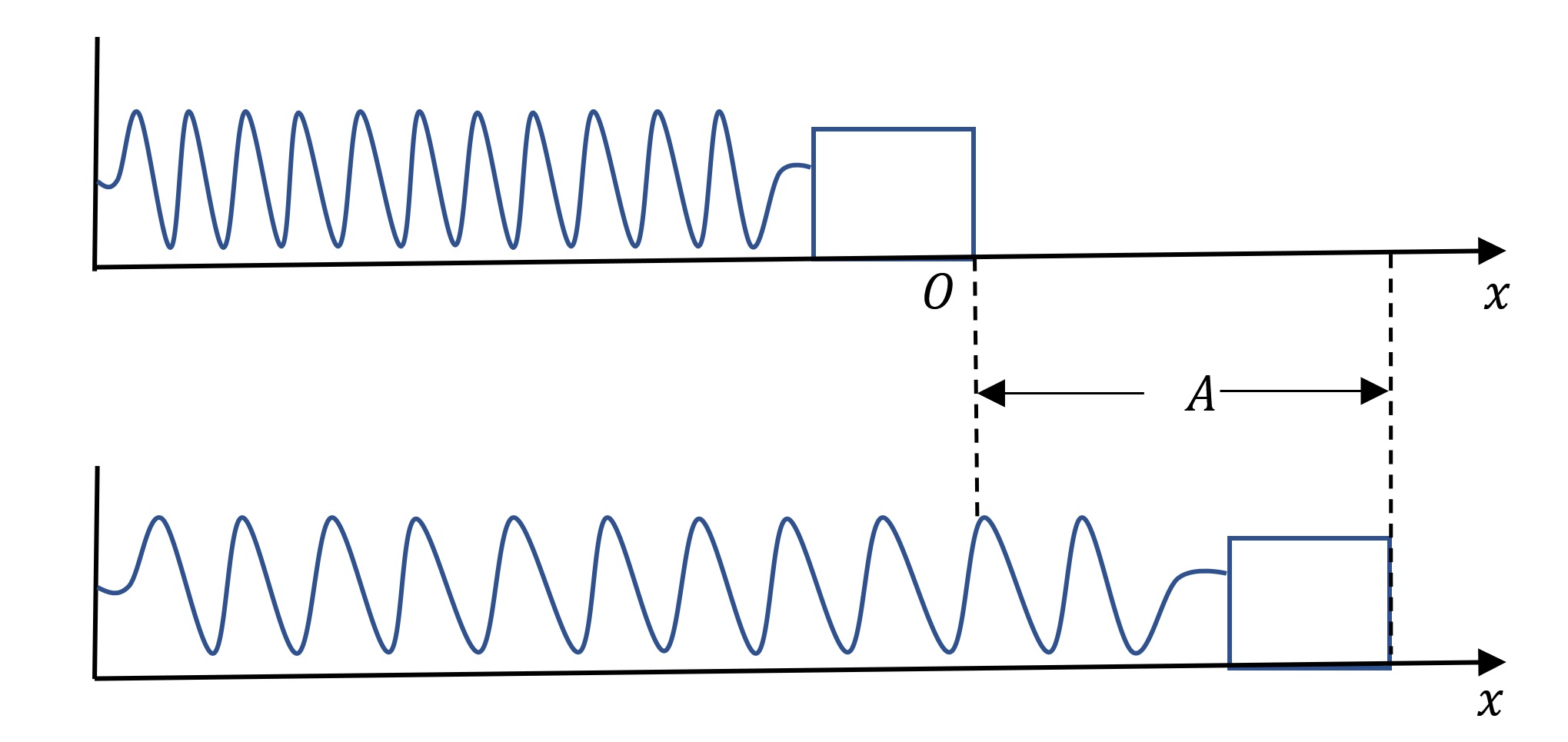}
\label{setup}
\caption{Setup of the Problem.}
\centering
\end{figure}
It is stretched to the right by distance $A$. It is then released from rest. We focus on the time for the motion from its initial position back to equilibrium, that is, the first quarter of a complete cycle. For later comparisons, we first give the exact value of one quarter of a period
\bea
T_{1/4}\,=\,\fft{\pi}{2}\sqrt{\fft{m}{k}},
\label{1/4}
\eea 
where the subscript indicates one quarter of a complete cycle. 
In Sec.~\ref{riemann} we will rediscover this value from our approach.

The first exercise is to find the time for the mass to move from the maximum displacement $x=A$ back to its equilibrium at $x=0$ under the \emph{average} of two forces exerted by the spring at these two positions. It will soon be seen that the notion of average, though seems primitive, may become quite powerful if executed appropriately.
The average force~$\ov F$ and the average acceleration $\ov a$ are 
\bea
\ov F\,=\,\fft12\,k\,A,\quad
\ov a\,=\,\fft{\ov F}{m}\,=\,\fft{k\,A}{2\,m}.
\eea
The kinematic formula for a motion with constant acceleration $a$, which  relates the displacement $x$, the initial velocity $v_0$, and the time $t$, is 
\bea
x\,=\,v_0\,t\,+\,\fft12\,a\,t^2.
\label{kinematic}
\eea
For our problem, the initial velocity is zero, the displacement is $A$, and the acceleration is~$\ov a$. Applying Eq. (\ref{kinematic}) gives the time, denoted as $t_{1/4}^{(1)}$, to move from the maximal displacement back to equilibrium, that is,
\bea
t_{1/4}^{(1)}\,=\,\sqrt{\fft{2\,A}{\ov a}}\,=\,2\sqrt{\fft{m}{k}}.
\eea
It is larger than one quarter of the exact period $T_{1/4}$ in Eq. (\ref{1/4}). The ratio between them is
\bea
r_1\,\equiv\,\fft{t_{1/4}^{(1)}}{T_{1/4}}\,=\,\fft{2}{\pi/2}\,=\,\fft{4}{\pi}\,\approx\,1.27. 
\label{r_1}
\eea

Now we divide the maximum displacement into two equal segments and  find the sum of time to cover each segment under the corresponding average force. We will find that the obtained approximate value is closer to the exact value $T_{1/4}$ in Eq. (\ref{1/4}).
In the first segment, the average force $\ov F_1$ and the average acceleration $\ov a_1$ are 
\bea
\ov F_1\,=\,\fft12\,(k\,A\,+\,\fft12\,k\,A)\,=\,\fft34\,k\,A,\quad
\ov a_1\,=\,\fft{3\,k\,A}{4\,m}.
\eea
Applying the kinematic formula in Eq. (\ref{kinematic}) gives the time $t_1$ to cover the first segment
\bea
t_1\,=\,\sqrt{\fft{2\,\times\,\fft12\,A}{\ov a_1}}
	\,=\,\fft{2}{\sqrt{3}}\,\sqrt{\fft{m}{k}}.
\eea
The velocity $v_1$ at time $t_1$ will be used to calculate the time for the second segment. It is obtained as
\bea
v_1\,=\,\ov a_1\,t_1
		\,=\,\fft{\sqrt{3}}{2}\,A\,\sqrt{\fft{m}{k}}.
\label{v_1}
\eea

In the second segment, the average force~$\ov F_2$ and the corresponding average acceleration~$\ov a_2$ are
\bea
\ov F_2\,=\,\fft12\,\times\,\fft12\,k\,A\,=\,\fft14\,k\,A,\quad
\ov a_2\,=\,\fft{k\,A}{4\,m}.
\eea
Applying the formula in Eq. (\ref{kinematic}) yields the equation determining the time $t_2$ covering the second segment, that is,
\bea
\fft12\,A\,=\,v_1\,t_2\,+\fft12\,\ov a_2\,t_2^2,
\eea
where $v_1$ is given in Eq. (\ref{v_1}).
The solution for $t_2$ is
\bea
t_2\,=\,2\,(2\,-\,\sqrt{3}\,)\sqrt{\fft{m}{k}}.
\eea

The total time $t_{1/4}^{(2)}$ to cover these two segments is
\bea
t_{1/4}^{(2)}\,=\,t_1\,+\,t_2
		\,=\,\fft{2}{\sqrt{3}}\,\sqrt{\fft{m}{k}}\,+\,2\,(2\,-\,\sqrt{3}\,)\sqrt{\fft{m}{k}}
		\,=\,\fft43(3-\sqrt{3})\sqrt{\fft{m}{k}},
\eea
where the superscript indicates that the whole displacement $A$ is divided into two equal segments.
The ratio between it and the exact value in Eq. (\ref{1/4}) is
\bea
r_2\,\equiv\,\fft{t_{1/4}^{(2)}}{T_{1/4}}\,=\,\fft{4\,(3-\sqrt{3})/3}{\pi/2}\,=\,1.08\,<\,1.27\,=\,r_1. 
\eea
It is closer to one quarter of the exact period. We expect that the difference from the exact value will decrease as the number of segments increases. In fact, as the number of segments goes to infinity, the exact value can be recovered. A full development along this direction is presented in the following section.


\section{Divide the Distance into Equal Segments: Riemann Sum}
\label{riemann}

The mass is released from rest at its maximal displacement. The goal is to find its position as a function of time~$t$, that is, $x(t)$. Moreover,  we denote the time for the mass to move from $x$ back to its equilibrium as~$\tau$. Clearly, the sum of $t$ and $\tau$ is equal to one quarter of a complete cycle denoted as $T'_{1/4}$, that is,
\bea
t+\tau\,=\,T'_{1/4}.
\label{t+tau}
\eea
The value of $T'_{1/4}$ is to be solved from our approach. In the end, we will find that $T'_{1/4}$ is indeed equal to $T_{1/4}$ given in Eq.~(\ref{1/4}). 
Figure~2 helps to illustrate the idea. 
\begin{figure}[h!]
\centering
\includegraphics[width=5.0in]{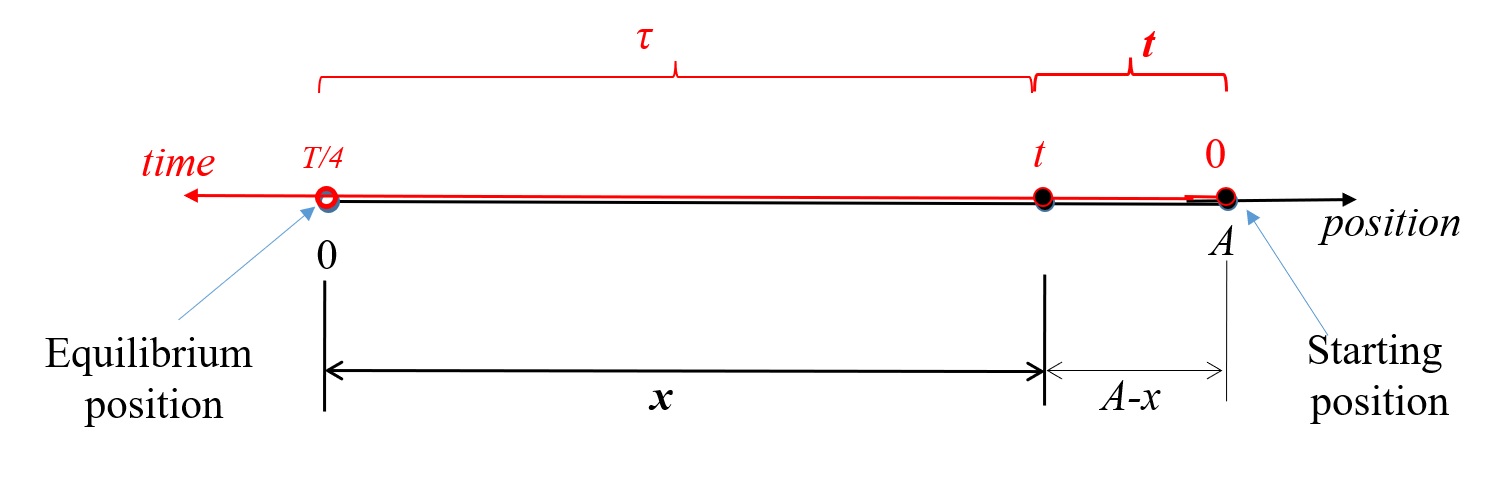}
\label{SHMfig1}
\caption{Illustration of the idea.}
\centering
\end{figure}

\begin{figure}[h!]
\centering
\includegraphics[width=5.0in]{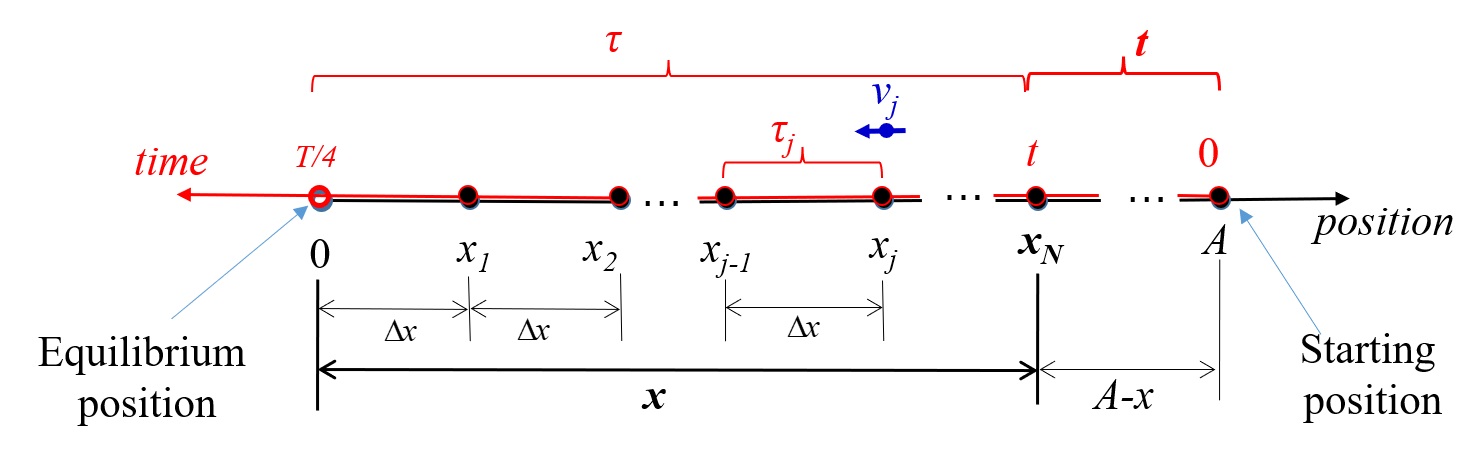}
\label{SHMfig2}
\caption{Divide the distance into $N$ equal segments.}
\centering
\end{figure}

The following is the key to our approach: The distance $x$ covered by the mass during the time interval $\tau$ is divided into $N$ equal segments. See Fig.~\ref{SHMfig2} for details. The motion in each segment is considered to have constant acceleration, under the \emph{average} of two forces exerted by the spring at the two ends of the segment under consideration. The quantity~$\tau$ can be obtained by calculating and summing over the time covering each segment. The sum appears to be the Riemann sum of a familiar integral which can be readily integrated, and the result is the same as that obtained from solving differential equation.

The size of each segment is denoted as $\D x=x/N$. The equilibrium is denoted as $x_0=0$ and the position of the right end of the $j$-th segment is $x_j=j\,\D x$. In particular, $x_N=x$. The time for the mass to cover the $j$-th segment is denoted as $\t_j$ and the total time $\tau$ to cover the distance $x$ is $\tau=\sum_{j=1}^N \t_j$.

By mechanical energy conservation, the \emph{speed} $v_j$ at position $x_j$ satisfies
\bea
\fft12\,m\,v_j^2\,+\,\fft12\,k\,x_j^2\,=\,\fft12\,k\,A^2,\quad j=0, 1, \dots,N.
\eea
Then $v_j$ is obtained as
\bea
v_j=\omega A\left(1-\d^2\,j^2\right)^{\fft12}, \quad \omega^2\equiv k/m,\quad \d\equiv\D x/A.\label{vj}
\eea
Recursively,
\bea
v_{j-1}=\omega A\left(1-\d^2\,(j-1)^2\right)^{\fft12}.\label{vjm1}
\eea
Both $v_j$ and $v_{j-1}$ together with the average acceleration found below will be used to find the time to cover the $j$-th segment.

In the $j$-th segment, the mass moves with an average acceleration under the average force on it. Applying Newton's second law and recalling $x_j=j\,\D x, \D x=x/N$ and $\d=\D x/A$ we have
\bea
m\ov {a}_j\,=\,\fft12\,k\,(x_{j-1}+x_j)\,=\,\fft12\,k\,\D x\,(2j-1)\,=\,\fft12\,k\,A\,\d\,(2j-1),\nn
\eea
obtaining
\bea
\ov {a}_j\,=\,\fft12\,\omega^2\,A\,\d\,(2j-1).\label{ajbar}
\eea
Note that $\omega\,A$ in Eq.~(\ref{vj}) and  $\omega^2\,A$ in Eq.~(\ref{ajbar}) are actually the maximum speed and acceleration for SHM, respectively. 

With Eqs.~(\ref{vj}),~(\ref{vjm1}), and~(\ref{ajbar}), we are ready to calculate the time for the mass to cover the $j$-th segment. Note that $v_{j-1}>v_j$, and $\t_j$ is obtained as
\bea
\t_j=\fft{v_{j-1}-v_j}{\ov{a}_j}=\fft{2}{\omega\,\d}\,\fft{\left(1-\d^2\,(j-1)^2\right)^{\fft12}-\left(1-\d^2\,j^2\right)^{\fft12}}{2j-1}\nn
\\=\fft{2\,\d}{\omega}\,\fft{1}{\left(1-\d^2\,(j-1)^2\right)^{\fft12}+\left(1-\d^2\,j^2\right)^{\fft12}}.\nn
\eea
Equivalently,
\bea
\omega\,A\,\t_j\,=\,\fft{2\,\D x}{\left[1-(x_{j-1}/A)^2\right]^{\fft12}+\left[1-(x_j/A)^2\right]^{\fft12}}.\label{tj}
\eea
Note the fact that $x_{j-1}<x_j$ (see Fig. \ref{SHMfig2}).
Then Eq. (\ref{tj}) gives 
\bea
\fft{\D x}{\left[1-(x_{j-1}/A)^2\right]^{\fft12}}<\omega\,A\,\t_j<\fft{\D x}{\left[1-(x_{j}/A)^2\right]^{\fft12}}.
\eea
Summing up $j$ from 1 to $N$ and recalling $\tau=\sum_{N=1}^{N} \tau_j$ give
\bea
\sum_{j=1}^N\fft{\D x}{\left[1-(x_{j-1}/A)^2\right]^{\fft12}}<\omega\,A\,\t<\sum_{j=1}^N\fft{\D x}{\left[1-(x_{j}/A)^2\right]^{\fft12}}.
\eea

The left sum is the so-called lower sum and the right the upper sum in Riemann integral~\cite{thomas, spivak}. Then, as $N$ goes to infinity, we have
\bea
\omega\,A\,\t=\int_0^x\fft{dx}{\sqrt{1-(x/A)^2}}=A\,\arcsin (x/A).
\eea
Equivalently,
\bea
x=A\sin \omega \t.\label{xtau}
\eea

Now we are ready to determine $T'_{1/4}$ in Eq.~(\ref{t+tau}). 
When $x=A$, the time for the mass to arrive at its equilibrium is just $T'_{1/4}$. Plugging them into Eq.~(\ref{xtau}) gives 
\bea
\sin \omega\, T'_{1/4}\,=\,1, \quad 
\omega\, T'_{1/4}\,=\,\pi/2,\quad 
T'_{1/4}\,=\,\fft{\pi}{2\,\omega}\,=\,\fft{\pi}{2}\,\sqrt{\fft{m}{k}},
\eea 
which is indeed $T_{1/4}$ in Eq.~(\ref{1/4}). 

According to Eq.~(\ref{t+tau}), $\t=T'_{1/4}-t$, and Eq.~(\ref{xtau}) becomes
\bea
x=A\sin \omega\,(T'_{1/4}-t)=A\,\sin \omega\,\left(\fft{\pi}{2\,\omega}-t\right)= A\,\cos \omega\,t,
\eea
which is exactly the same as that obtained from solving the differential equation for SHM~\cite{morin, thomas}.

\section{Divide the Time into Equal Intervals: Recurrence Relation} \label{recursive}
\begin{figure}[h!]
\centering
\includegraphics[width=5.0in]{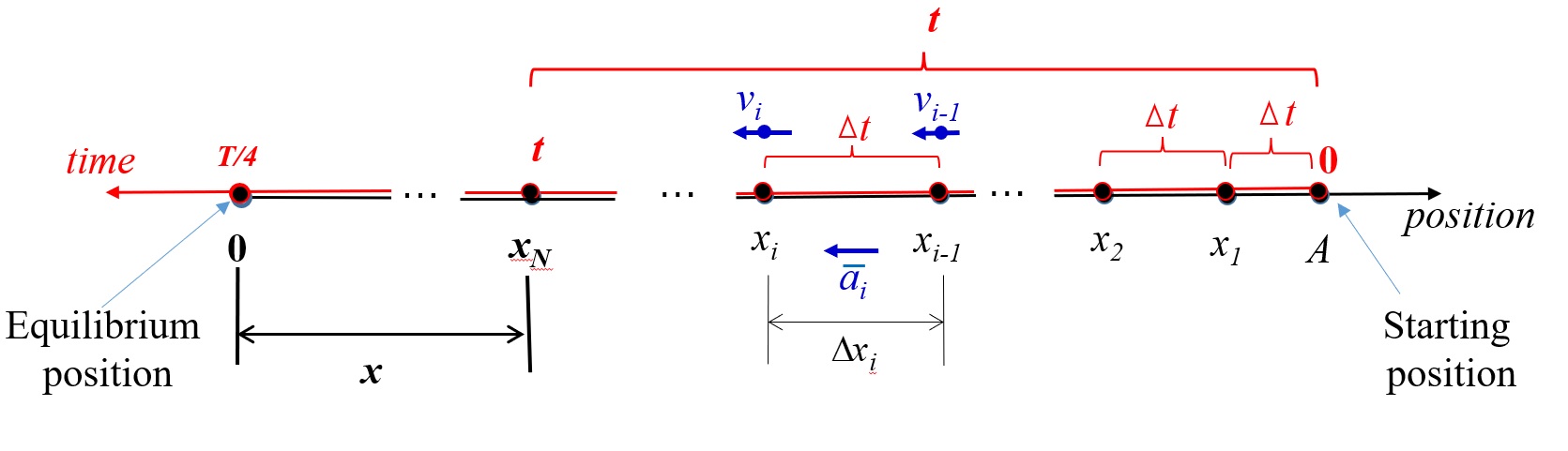}
\label{SHMfig3}
\caption{Divide the time into $N$ equal intervals.}
\centering
\end{figure}

Another approach is to divide the time $t$ after the mass is released into $N$ equal intervals. See Fig.~\ref{SHMfig3} for details.
A recurrence relation for the position of the mass will be obtained. In the end, the large-$N$ limit of its solution results in the same one as obtained from solving differential equations.

To be specific, let $x_i$ and $v_i$ denote the position and speed of the mass at the end of the~$i$-th time interval, respectively. The size of each time interval is $\D t=t/N$, and the distance covered in the $i$-th time interval is denoted as $\D x_i=x_{i-1}-x_i$, where~$x_i$ is the position of the mass at the end of the $i$-th time interval. Note that $x_0=A$, $x_i<x_{i-1}$, and $v_i>v_{i-1}$. The average speed during the $i$-th interval is  
\bea
\fft12\,(v_i+v_{i-1})=\fft{\D x_i}{\D t}.    \label{kine1}
\eea

The key step in our approach is the following. Within each time interval, the motion is considered to have  constant acceleration, under the average of two elastic forces with spring extensions corresponding to the two ends of the time interval, that is,
\bea
v_i-v_{i-1}=\overline{a}_i\,\D t,     \label{kine2}
\eea
where $\overline{a}_i$ is the average acceleration in the $i$-th time interval. It is obtained by applying Newton's second law, that is,
\bea
m\ov a_i\,=\,\fft12\,k\,(x_i+x_{i-1}),\quad
\overline{a}_i=\fft12\,\omega^2\,(x_i+x_{i-1}), \quad  \omega^2\equiv k/m.\label{abar}
\eea
Solving $v_i$ and $v_{i-1}$ from Eqs. (\ref{kine1}) and (\ref{kine2}) gives
\bea
2v_i=2\,\fft{\D x_i}{\D t}+\overline{a}_i\,\D t, \label{vi}\\ 
2v_{i-1}=2\,\fft{\D x_i}{\D t}-\overline{a}_i\,\D t. \label{vim1}
\eea

Recursively, from Eq. (\ref{vim1}) we have an alternative expression for $v_i$, that is,
\bea
2v_i=2\,\fft{\D x_{i+1}}{\D t}-\overline{a}_{i+1}\,\D t. \label{vip}
\eea
Equating Eq. (\ref{vi}) with Eq. (\ref{vip}) gives
\bea
\D x_{i+1}-\D x_i=\fft12\,(\overline{a}_i+\overline{a}_{i+1})\,\D t^2.
\eea
With the expression for $\ov a_i$ in Eq. (\ref{abar}), a recurrence relation for $x_i$ is obtained as 
\bea
(1+\a^2)\,x_{i+1}-2\,(1-\a^2)\,x_i+(1+\a^2)\,x_{i-1}=0, \quad \a\equiv \fft12\,\omega\D t.\label{master}
\eea

To solve the recurrence relation in Eq. (\ref{master}), we need two initial conditions. One is obviously the initial position of the mass, that is, $x_0=A$. The other can be found as follows. The distance $x_1$ covered in the first time interval is 
\bea
A-x_1=\fft12\,\overline{a}_1\,\D t^2=\fft14\,\omega^2(A+x_1)\,\D t^2=\a^2\,(A+x_1),
\eea 
where the second equality follows from Eq.~(\ref{abar}) and the third from the definition of~$\a$ in Eq.~(\ref{master}). Then the two initial conditions are 
\bea
x_0=A,\quad x_1=\fft{1-\a^2}{1+\a^2}\,A. 
\label{ini}
\eea

Equation (\ref{master}) with initial conditions in Eq. (\ref{ini}) can be solved by writing down
the characteristic equation \cite{epp}, that is,
\bea
(1+\a^2)\,r^2-2\,(1-\a^2)\,r+(1+\a^2)=0.\label{chara}
\eea
Its two roots are
\bea
r_+=\fft{1+i\,\a}{1-i\,\a},\quad r_-=\fft{1-i\,\a}{1+i\,\a},\quad i^2=-1.
\eea
The general solution to the recurrence relation is 
\bea
x_k=\lambda_+\,r_+^k\,+\, \lambda_-\,r_-^k,\quad k=0,1,\cdots,N,
\label{general}
\eea
where $\lambda_+$ and $\lambda_-$ are constants to be determined by initial conditions in Eq.~(\ref{ini}).
Substituting initial conditions in Eq. (\ref{ini}) into Eq. (\ref{general}) gives
\bea
\lambda_+\,=\,\lambda_-\,=\,\fft12\,A.
\eea
Finally, we obtain the solution to the recurrence relation in Eq.~(\ref{master}) with initial conditions in Eq.~(\ref{ini}) as
\bea
x_k=\fft12\,A\,(r_+^k\,+\, r_-^k),\quad k=0,\cdots,N.
\eea
In particular, the position of the mass at time $t$, or equivalently, at the end of the $N$-th time interval is 
\bea
x_N=\fft12\,A\,(r_+^N\,+\, r_-^N).\label{xN}
\eea

The solution to the SHM can be obtained by taking the large-$N$ limit of Eq.~(\ref{xN}). For this purpose, we write 
\bea
r_+^N=\left(\fft{1+i\,\a}{1-i\,\a}\right)^N=\fft{\left(1+\fft12\,i\,\omega\,\D t\right)^N}{\left(1-\fft12\,i\,\omega\,\D t\right)^N}=\fft{\left(1+\fft12\,i\,\omega\,t\,\fft1N\right)^N}{\left(1-\fft12\,i\,\omega\,t\,\fft1N\right)^N}.
\eea\nn
In the large-$N$ limit, the numerator becomes $e^{i\,\omega \,t/2}$ and the denominator $e^{-i\,\omega \,t/2}$. Thereby, the limit for $r_+^N$ is $\lim_{N \to \infty} r_+^N=e^{i\,\omega \,t}$ and similarly $\lim_{N \to \infty} r_-^N=e^{-i\,\omega \,t}$.
Finally, the large-$N$ limit of Eq. (\ref{xN}) is
\bea
x(t)=\lim_{N \to \infty} x_N=\fft12\,A\,\left(e^{i\,\omega \,t}\,+\,e^{-i\,\omega \,t}\right)=A\,\cos \omega\,t,\label{xt}
\eea
which is exactly the same as that obtained from solving differential equation for SHM~\cite{morin, thomas}.

\section{Conclusion and Discussion}\label{con}
This paper presents two novel approaches to solve the classic problem of SHM. 
The key idea is dividing the distance or the time into $N$ equal segments or intervals, and the motion is approximated as one with constant acceleration. The large-$N$ limit is taken to obtain exactly the same result as that from solving differential equations.

The value of our approaches lies in two aspects. On the one hand, they enable students to attack a problem at an early stage using elementary physics and mathematics rather than after they have learned enough calculus or even the theory of differential equation at a much later stage. On the other hand, even this paper is presented to students after they are able to solve differential equations, they can still learn quite some mathematics and problem solving skills from this paper. More importantly, by comparing various approaches, they have a good chance to further appreciate the power, simplicity, and elegance of calculus.

\end{document}